\documentclass[11pt]{article}

\usepackage[T1]{fontenc}
\usepackage{lmodern}
\usepackage[letterpaper,margin=1in]{geometry}
\usepackage{amsmath}
\usepackage[round]{natbib}
\usepackage{booktabs}
\usepackage{graphicx}
\usepackage{tabularx}
\usepackage{array}
\usepackage{multirow}
\usepackage{placeins}
\usepackage{xspace}
\usepackage{microtype}
\usepackage{url}
\usepackage{fancyhdr}
\usepackage[colorlinks=true,allcolors=black]{hyperref}

\newcommand{\Items}{\mathcal{I}}

\newcommand{\prefix}{\operatorname{pref}}

\providecommand{\Description}[1]{}

\newcolumntype{Y}{>{\raggedright\arraybackslash}X}

\newcommand{\tool}{SIDScope\xspace}

\newcommand{\preprintfooter}{\footnotesize Author-prepared preprint.}
\fancypagestyle{plain}{%
  \fancyhf{}%
  \fancyfoot[L]{\preprintfooter}%
  \fancyfoot[R]{\footnotesize\thepage}%
  }

\title{\textbf{SIDScope: A Diagnostic Resource for Semantic-ID Interfaces in
Generative Recommendation}\thanks{Resource entry point:
\href{https://github.com/jdding/SIDscope/releases/tag/v2.0.0}{SIDScope 2.0.0 release}.}}

\author{%
  Jiandong Ding\thanks{Corresponding author: \texttt{dingjiandong2@huawei.com}},
  Huijie Qin, Tiandeng Wu, Yi Cao\\[4pt]
  Huawei Technologies Co., Ltd., Shanghai, China\\[2pt]
  \texttt{\{dingjiandong2, qinhuijie, wutiandeng1, caoyi23\}@huawei.com}}

\date{October 9, 2026}

\begin{document}
\maketitle
\begin{abstract}

Semantic-ID tokenizers assign code sequences to catalog items, which generative
recommenders predict and resolve back to items. Reusing a released mapping
requires checking its code organization, item resolution, and compatibility
with an existing model after a refresh. SIDScope provides these checks through
source-traced adapters, mapping diagnostics, candidate-retrieval tests, and
generated-trace analysis. Its inventory contains nine exports from seven
tokenizer families across Amazon and Yelp, including eight executable routes
and one auditable snapshot. The analyses establish three distinctions. First,
code allocation and item addressability differ: in GRID, tail items occupy
95.7 distinct codes per 100 items, yet only 82.2\% have globally unique codes.
Second, prefix alignment calibrates retrieval operations that use prefixes.
With fixed candidate budgets and ranking, original assignments on Beauty and
Instruments expose more targets than controls preserving code capacity and
coarse popularity. Resolving identical generated paths also shows that an
explicit item-selection rule recovers part of the gap between target-path
survival and singleton credit. Third, a released refresh repairs missing
codes, while model adaptation separately restores new-item recall. SIDScope
supplies code, diagnostic summaries, and verification tools to compare
mappings, evaluate their use in retrieval and generation, and test model reuse
after changes.

\end{abstract}

\begin{center}
\begin{minipage}{0.94\textwidth}
\small
\textbf{Prior-publication disclosure.}
This article is a substantial extension of the CIKM 2026 Resources
paper, \emph{SIDInspector: A Mapping-First Diagnostic Resource for Semantic-ID
Tokenizers}~\citep{sidinspector2026mapping}. Table~\ref{tab:v0-delta}
distinguishes its inherited foundations from the scientific evidence and
resource capabilities introduced here.
\end{minipage}
\end{center}

\section{Introduction}

Semantic-ID tokenizers give catalog items discrete code sequences, which
generative recommenders predict as paths through an item address space
\citep{rajput2023tiger,ju2025grid}. Once exported, this mapping can be reused
with a different generator or refreshed as the catalog changes. Its quality
therefore matters beyond the ranking score of the model that first used it.

Recent tokenizers learn this address space through quantization,
collaborative objectives, collision-aware assignment, and variable-length
representations
\citep{wang2024letter,zhu2024cost,zheng2023lcrec,liang2026resid,wei2026card,cheng2026capsid}.
A released mapping, however, leaves practical questions unanswered. Do its
prefixes retain behavioral neighbors? Can a generated leaf identify one item,
or does it need a resolver? Does repairing catalog coverage preserve the
addresses an existing generator has learned? Recall and NDCG summarize a
complete recommendation pipeline; they do not isolate these interface
properties.

SIDScope is a diagnostic resource for investigating three research questions:
\begin{enumerate}
 \item \textbf{What differs across mappings?} We measure code use, shared
 addresses, behavioral alignment, popularity allocation, and trie structure.
 These properties are non-interchangeable: unique full codes can coexist with
 weak prefix alignment, and distinct codes within a popularity stratum need
 not identify its items uniquely across the catalog.
 \item \textbf{When do those differences affect downstream outcomes?}
 We connect prefix alignment to candidate exposure, separate construction
 calibration from held-out and non-prefix checks, and distinguish a generated
 path from its resolved item. Fixed-budget assignment controls and frozen-beam
 resolver comparisons test concrete uses of the mapping.
 \item \textbf{What must change after a mapping refresh?}
 We follow a coverage repair through a paired mapping audit and a predeclared
 generator-handoff decision, separating a valid new mapping from a model ready
 to use it.
\end{enumerate}

The resource builds on SIDInspector's mapping-first diagnostics
\citep{sidinspector2026mapping}. It adds source-traced adapters, consumer-facing
evaluation protocols, generated-trace accounting, and paired refresh evidence.
A common contract joins mappings to their catalogs, interactions, and optional
traces; it supplies the measurement foundation for the three questions rather
than an additional quality score.

SIDScope supplies adapters, measurements, and reproducible evidence for nine
named exports from seven families across Amazon and Yelp. Eight exports have
executable conformance reports; one is an auditable snapshot. Using this
resource, we establish three findings. First, code allocation within a
popularity group and unique item addresses across the catalog can differ
substantially, so both are needed to assess tail-item resolution. Second,
mapping structure matters through the operation that uses it. Fixed-budget
tests on Beauty and Instruments isolate useful prefix assignments beyond
code capacity and coarse popularity; comparisons on identical generated
paths show how an explicit resolver converts shared codes into item hits.
Third, the released DACT case separates coverage repair from checkpoint
reuse: repairing missing codes leaves new-item recall at zero until the model
is adapted. Matched controls, disjoint evaluation users, uncertainty checks
at the artifact and catalog units, and a predeclared handoff gate support
these conclusions within the evaluated settings.

\begin{figure*}[t]
  \centering
  \includegraphics[width=\textwidth]{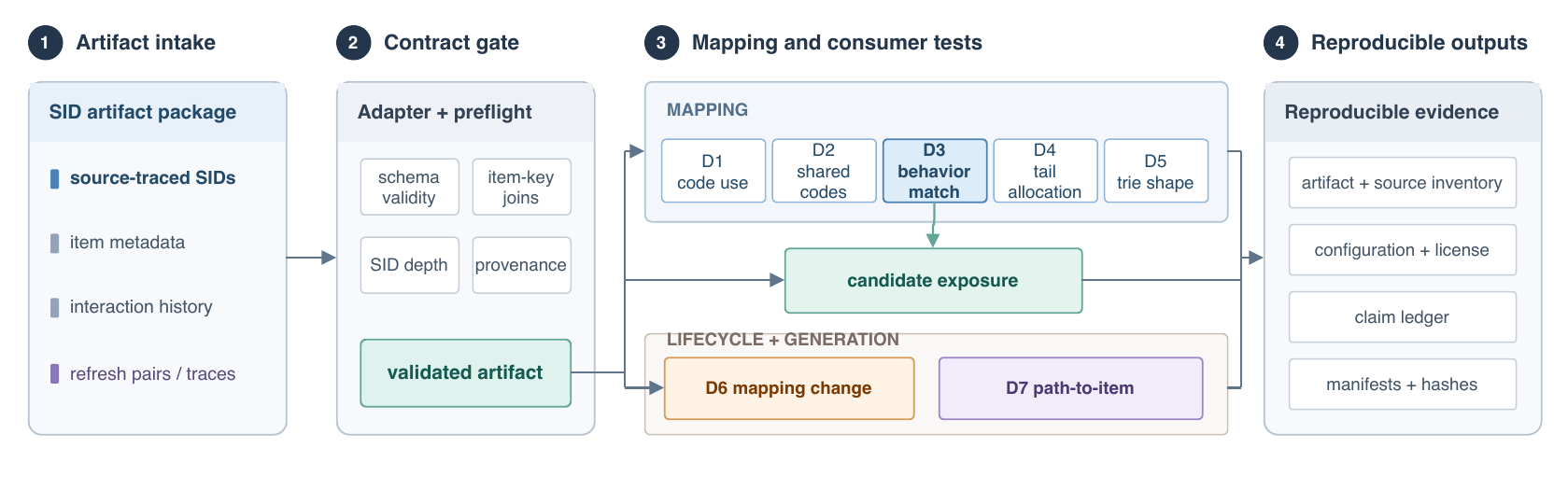}
  \caption{SIDScope connects source-traced inputs to mapping diagnostics,
  retrieval and generation tests, and reproducible outputs. D1 measures
  code-space use, D2 shared-address risk, D3 prefix--behavior alignment,
  D4 popularity allocation, D5 trie structure, D6 mapping change, and
  D7 path-to-item resolution. Source, configuration, and license records bind
  these measurements to the mapping being tested.}
  \Description{A four-stage evidence-chain schematic. A SID artifact package
  containing an item-to-SID mapping, metadata, interactions, and optional
  refresh or trace inputs passes schema, join, depth, and provenance checks.
  The validated artifact is then profiled by D1-D5 mapping diagnostics,
  candidate exposure, D6 mapping-change pairs, and D7 path-to-item accounting. The final
  resource binds the artifact inventory to source, configuration, license, claim,
  manifest, and hash records.}
  \label{fig:evidence-chain}
\end{figure*}

Figure~\ref{fig:evidence-chain} shows how the resource connects these questions
to one inspectable artifact record.

Table~\ref{tab:v0-delta} makes the extension over SIDInspector V0
self-contained by linking each added capability to its article evidence.

\begin{table}[!htbp]
  \centering
  \caption{Scientific and resource extension beyond SIDInspector V0. The final
  column points to the evidence introduced and evaluated in this article.}
  \label{tab:v0-delta}
  \small
  \setlength{\tabcolsep}{3.2pt}
  \renewcommand{\arraystretch}{1.04}
  \begin{tabularx}{\textwidth}{@{}>{\raggedright\arraybackslash}p{0.18\textwidth}>{\raggedright\arraybackslash}p{0.22\textwidth}>{\raggedright\arraybackslash}p{0.36\textwidth}Y@{}}
    \toprule
    Question & SIDInspector V0 & SIDScope extension & Where shown \\
    \midrule
    What is inspected? & Static item-to-code mappings & Artifact states spanning mappings, exposure, traces, and package records & Figure~\ref{fig:evidence-chain}; Sections 3 and 7 \\
    Which artifacts enter? & Reconstructed controls and snapshots & Source-traced named routes separated from stress and control rows & Tables~\ref{tab:artifact-coverage} and \ref{tab:adapter-conformance} \\
    What can D3 support? & D1--D5 diagnostic tables & Prefix calibration, fixed-budget assignment controls, and measurement checks & Tables~\ref{tab:candidate-exposure}, \ref{tab:fixed-budget}, and \ref{tab:measurement-audit} \\
    What do traces add? & D7 input hook & Constraint-aware labels and resolver comparisons on identical generated paths & Tables~\ref{tab:g20-trained-traces} and \ref{tab:resolver-policy}; Figure~\ref{fig:d7-addressability-gap} \\
    How broad is the matrix? & Musical worked example & Nine named exports, seven families, two data ecosystems, controls, and trained traces & Tables~\ref{tab:artifact-coverage}, \ref{tab:diagnostic-profile}, and \ref{tab:g20-trained-traces} \\
    What changes over time? & Mapping snapshots & Paired D6 refresh audit, mapping repair, and generator handoff & Table~\ref{tab:dact-handoff} \\
    How is it reproduced? & Toolkit-style release & C0--C5 conformance, frozen evidence summaries, hashes, and quickstart & Tables~\ref{tab:artifact-coverage}, \ref{tab:adapter-conformance}, and \ref{tab:resource-entrypoint} \\
    \bottomrule
  \end{tabularx}
\end{table}

Section 2 positions the resource. Section 3 defines its inputs and
measurements. Sections 4--6 address mapping differences, downstream
consequences, and refresh decisions, respectively. Sections 7--8 discuss
interpretation and reuse; the appendices document protocols and measurement
checks.

\section{Related Work}

\paragraph{Semantic-ID construction and use.}
Generative retrieval first made identifiers themselves part of the learned
retrieval interface, including learned document IDs, substring identifiers, and
neural corpus indexes \citep{tay2022dsi,bevilacqua2022seal,wang2022nci}. In
recommendation, text-to-text and sequence-transduction formulations likewise
turn recommendation into generation over tokens or actions
\citep{geng2022p5,zhai2024hstu,zhang2025instruction}. TIGER-style generative retrieval introduced
the now-common pattern of assigning items to discrete semantic code sequences
and training a sequence model to emit those codes as recommendations
\citep{rajput2023tiger}. Related work has also studied how item identifiers
should be indexed for recommendation foundation models and how semantic IDs can
improve generalization in ranking or retrieval settings
\citep{hua2023indexids,singh2023bettersemanticids,chen2025idllm}. Surveys of generative
search and recommendation place these systems in a broader shift from
nearest-neighbor retrieval toward sequence-generation interfaces
\citep{li2024llmgrsurvey,li2025generativeir,hou2025tokenizationperspective}.

The dominant SID construction path inherits discrete representation learning
from VQ-VAE and residual quantization from RQ-VAE
\citep{oord2017vqvae,lee2022rqvae}. Subsequent work changes three parts of this
construction. First, content, interaction, and multimodal supervision reshape
what the codes preserve
\citep{zhu2024cost,wang2024letter,qu2025tokenrec,zhai2025mql4grec,luo2025qarm,zheng2026utgrec,wang2024eager,yang2025cobragrec,chen2026syngr,qiao2026textasvision,xiao2026unger}.
Second, recommender-native, differentiable, dynamic, and end-to-end objectives
move assignment closer to the downstream model
\citep{liu2024elit,fu2026diger,liang2026resid,wang2026pit,jiang2026unisid}.
Third, binary and tree-structured designs make identifier topology explicit, while
joint, transferable, and adaptive schemes broaden the contexts in which an SID
must remain useful
\citep{zheng2026topogr,penha2025jointsid,liu2025discrec,hu2025gencdr,wang2026sa2crq,hou2026expressiveness}.
Time-aware systems further incorporate temporal context or elapsed gaps into
the identifier pipeline \citep{nian2026chronoid,huang2026chronosid}.
Collision-aware and adaptive-overlap methods distinguish harmful aliasing from
admissible sharing
\citep{hu2026quasid,pan2026adasid}, while non-uniform quantization and
variable-length designs alter how item mass and semantic resolution are
allocated across code paths
\citep{wei2026card,cheng2026capsid,xia2026acerec,khrylchenko2026variablesid,wang2026varlenrec,zhang2026hgrec,huang2026asymrec,yan2026aktrec}.
SID artifacts also change across time, domains, and deployment settings.
Staleness and refresh work studies how collaborative SIDs drift under new logs
\citep{feng2026dact,baikalov2026staleness}, while dynamic or context-specific
assignment systems refine early mappings, live-streaming identifiers, or
category-guided prefix tries
\citep{guan2026dream,shi2026ssrlive,zhang2026calir}. Industrial SID work reports
use cases in ranking, short-video search, large-scale retrieval, and
LLM-enhanced recommendation trade-offs
\citep{ju2026snapchatsid,li2026sidcoord,xu2026largescalegenrec,li2026taiji}.
Recent industrial user-modeling work also extends SID vocabularies from item
address spaces to SID-based user tokens, making generated-token diversity,
cross-scenario conditioning, query integration, and representation refresh part
of the SID interface surface \citep{liu2026tokenminds}.
These designs motivate inspecting the exported mapping as a reusable address
space: its code assignment can be measured separately from the model that
created or consumes it.

\paragraph{Resources for generative recommendation.}
OpenP5 packages data processing, item indexing, model training, and evaluation
for LLM-based recommendation \citep{xu2024openp5}. GRID organizes modular
generative-recommendation components and controlled design comparisons
\citep{ju2025grid}. More general toolkits, including LensKit, RecBole, and
Elliot, support reproducible recommendation experiments
\citep{ekstrand2018lenskit,zhao2021recbole,anelli2021elliot}. These resources
organize model development and evaluation. SIDScope instead inspects the
mapping that can be reused across models: source-traced adapters make exports
comparable, and consumer-specific tests connect their measured properties to
retrieval, item resolution, and refresh decisions.

\paragraph{Mapping and consumer evaluation.}
Recommendation evaluation has long required more than a single leaderboard
metric, from classic collaborative-filtering methodology and sampled-metric
validity to later audits of reproducibility and progress
\citep{herlocker2004evaluating,krichene2020sampled,dacrema2021troubling}.
Recent studies identify artifacts and protocol details needed to reproduce
results, including in LLM-based recommendation
\citep{lops2024p5repro,shehzad2025code}. RecList makes recommendation behavior
inspectable through behavioral tests \citep{chia2022reclist}. In generative
recommendation, SID information bottlenecks, finite-beam pruning, and ranking
objectives all affect the relation between identifiers and recommendation
quality \citep{liu2026sidscaling,zeng2024planning,tang2024listwise,zhou2024roger}.

Mapping-first diagnostics expose the address space used by a generator
\citep{sidinspector2026mapping}, while collision-aware evaluation distinguishes
SID matching from item credit \citep{zhang2026sidreliability}.
Closed-loop simulation and temporal reachability studies further show that
feedback and time change which code paths are exposed or reachable
\citep{yang2026recloop,peng2026colditems}. These results motivate a reusable
record of the mapping, its source, and its consumer.

SIDScope connects this mapping record to explicit consumers. Fixed-budget
assignment controls test candidate retrieval, frozen-beam resolver comparisons
test item selection, and paired refreshes test whether a checkpoint can use
changed codes. The resource complements model-level evaluation by making
these intermediate decisions reproducible.

\section{The Resource and Its Measurements}\label{sec:resource-design}

\subsection{An Item-to-Code Mapping as a Reusable Interface}

\begin{figure*}[!t]
  \centering
  \includegraphics[width=\textwidth]{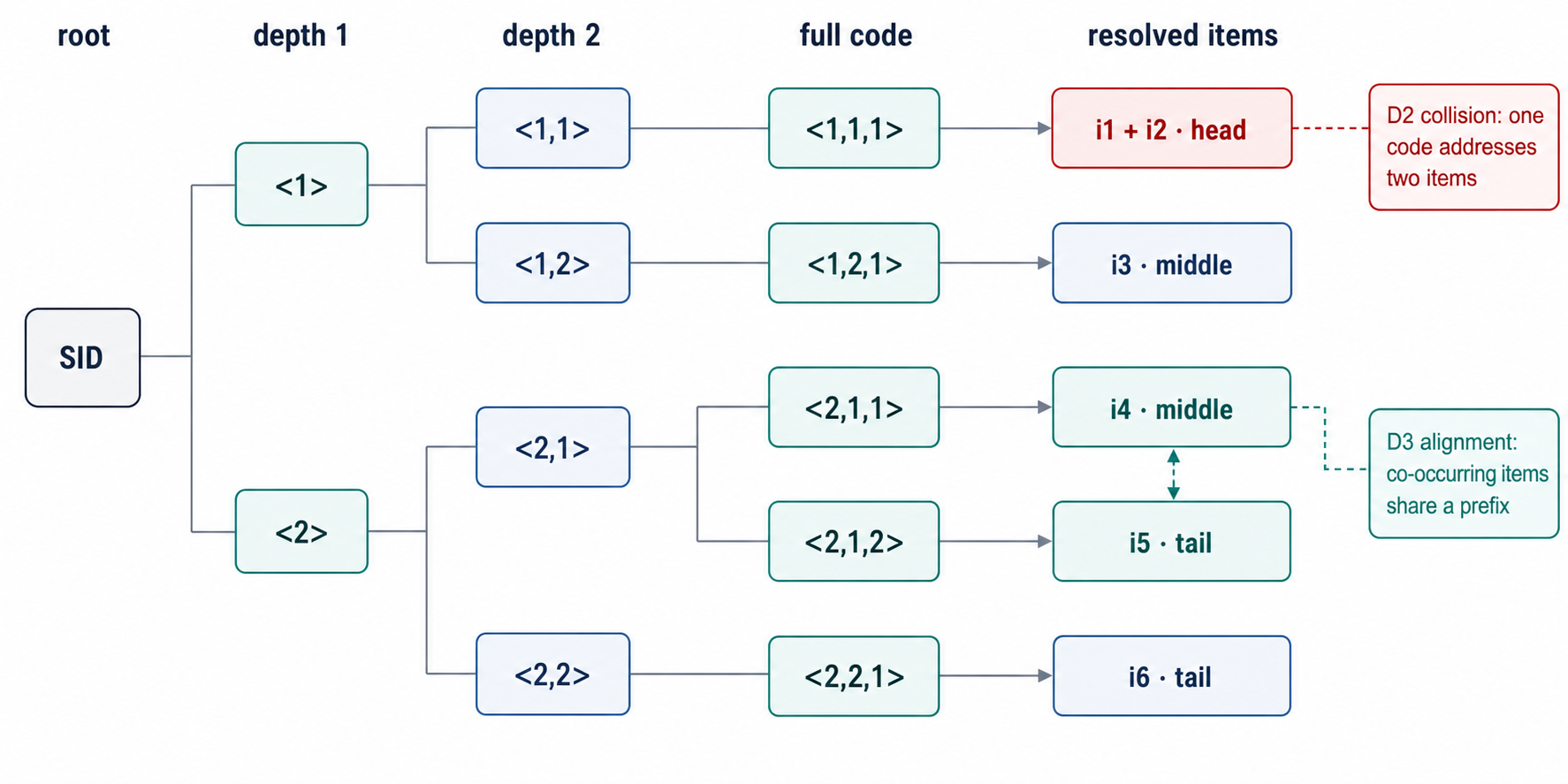}
  \caption{One mapping, distinct diagnostic questions. Items $i_1$ and $i_2$
  share full code $\langle1,1,1\rangle$, creating a D2 collision. Items $i_4$
  and $i_5$ co-occur and share prefix $\langle2,1\rangle$ but have different
  leaves, illustrating the relation measured by D3. All values are illustrative.}
  \Description{A three-level SID tree connects six items to five occupied full
  codes. Two items share a full code, while two co-occurring items share a
  prefix and retain distinct leaves.}
  \label{fig:sid-mapping-example}
\end{figure*}

Let $\Items$ be a catalog and $c(i)=(z_1,\ldots,z_L)$ the Semantic ID of item
$i$. The full sequence is a leaf address; its prefix
$\prefix_k(c(i))=(z_1,\ldots,z_k)$ identifies a region of the code trie.
Different items may share a leaf, and different leaves may share a useful
prefix. Figure~\ref{fig:sid-mapping-example} illustrates why these properties
require separate measurements.

SIDScope records the mapping together with its source identity, catalog,
configuration, and reuse terms. Adapters convert upstream files or checkpoints
into a common item-to-code table. Metadata and interactions support semantic
and popularity analyses; paired mappings support refresh analysis; generated
beams support path-to-item accounting. These inputs can be added to the same
artifact record without retraining the tokenizer.

The resource contains adapter and diagnostic code, normalized evidence
summaries, source manifests, conformance reports, table snapshots, figure
data, and CPU verifiers. Upstream checkpoints and data remain at their source
when redistribution is restricted. A researcher can therefore inspect a
mapping on its own, then attach consumer-specific evidence as it becomes
available. Source and replay details are collected in
Appendix~\ref{sec:protocol-details}.

\subsection{Mapping Structure: D1--D5}

\paragraph{Code use and shared addresses.}
D1 reports the number of occupied symbols, base-2 entropy, and Gini of the
symbol-frequency distribution at each code level. These are marginal
distributions, not counts of joint prefixes; D5 records the latter.
Let $m_s$ be the number of catalog items with full code $s$. D2 is the
collision-item rate,
\[
 D2=\frac{\sum_s m_s\mathbf{1}[m_s>1]}{|\Items|}.
\]
It counts every item sharing a full address. In contrast, the duplicate-SID
rate $1-|\{s:m_s>0\}|/|\Items|$ measures the shortfall in distinct leaves.
For example, a shared two-item leaf contributes two affected items to D2 but
one missing leaf to the shortfall. Prefix collision rates use the same item
denominator at the selected depth.

\paragraph{Behavioral alignment.}
D3 asks what fraction of observed item-neighbor relations a prefix preserves.
Its protocol $\pi=(m,U,B)$ selects users with between two and $U$ distinct
mapped items, accumulates at most $B$ unordered co-occurrence pair events,
and retains up to $m$ neighbors per item. Users exceeding $U$ are excluded
from this bounded graph construction, not truncated. Co-occurrence counts
select the top neighbors. The reported statistic then gives each retained
directed edge equal weight:
\[
 D3(k;\pi)=\frac{1}{|E_\pi|}
 \sum_{(i,j)\in E_\pi}
 \mathbf{1}[\prefix_k(c(i))=\prefix_k(c(j))].
\]
We call this \emph{edge-micro prefix-neighbor recall}. It is distinct from
averaging per-item recalls or weighting retained edges by co-occurrence
frequency. Appendix~\ref{sec:measurement-audit} compares all three estimands
against the executable measurements.

Interaction-derived diagnostics use the declared training split, or the
provided reference interactions when no split column exists. User--item
duplicates are removed before graph construction. The compact profile table
uses $\pi_{\mathrm{C3}}=(5,50,10{,}000)$; the public CLI default is
$\pi_{\mathrm{CLI}}=(20,200,2{,}000{,}000)$. The LETTER and LC-Rec C3 graphs
consume 9,817 pair events. Each downstream analysis declares its own instance,
including item ordering and tie handling. Thus D3 compares mappings under a
specified neighborhood construction rather than supplying a protocol-free
quality score.

\paragraph{Popularity allocation and global resolution.}
D4 divides the catalog into equal-count popularity tertiles using reference
training-event counts and deterministic item order for ties. For stratum $b$,
its allocation ratio is
\[
 A_b=\frac{|\{c(i):i\in b\}|}{|b|}.
\]
This describes how many distinct codes the stratum occupies. To answer whether
its items are individually addressable in the \emph{whole catalog}, we also
report
\[
 S_b=\frac{|\{i\in b:m_{c(i)}=1\}|}{|b|}.
\]
A tail item may occupy a distinct code within the tail yet share it with a
head item. $A_b$ and $S_b$ therefore answer different questions;
Section~\ref{sec:mapping-differences} measures their separation.

D5 records SID depth, occupied joint prefixes, full-code leaves, and fan-out.
These counts describe the structure a decoder traverses. Timing requires a
specified generator, implementation, hardware, and batch policy. The current
matrix uses fixed-depth exports; a variable-length export must declare
termination and be analyzed by length stratum.

\subsection{Consumer Outcomes and Mapping Changes}

D6 compares paired mappings over stable item IDs, recording coverage changes
and code churn. A revised mapping can restore coverage while changing the
addresses learned by an existing generator. Section~\ref{sec:resource-use}
tests the resulting reuse decision on a released mapping--checkpoint pair.

D7 joins generated paths to the canonical reverse mapping. We distinguish a
valid path, a path containing the target among its resolved items, and a path
that identifies the target uniquely. \emph{Aliasing} is the mapping property
of shared full codes; an \emph{ambiguous path} is a generated leaf resolving to
multiple items. A deployment may resolve that ambiguity with an additional
selection rule. We evaluate such rules on the same frozen beams in
Section~\ref{sec:resolver-decision}, keeping path order and cutoff fixed.

These measurements are non-interchangeable: code allocation does not establish
global addressability, and path survival does not determine a resolver's item
output. They guide the next comparison while preserving the distinction
between a mapping property and a consumer outcome.

\subsection{Coverage and Comparable Evidence}

An artifact route identifies a method family, catalog, and export branch.
Table~\ref{tab:artifact-coverage} lists nine source-traced named exports from
seven families across Amazon and Yelp. Eight have executable C0--C5 reports.
The ninth, ReSID/Musical, is an auditable snapshot, not an executable route;
it contributes to source coverage but not to the eight-row diagnostic matrix.
Conformance checks establish source identity, schema coherence, joins,
bounded execution, replay identity, and inventory consistency
(Table~\ref{tab:adapter-conformance}).

\begin{table}[!htbp]
  \centering
  \caption{Artifact inventory. Eight executable tokenizer routes pass C0--C5;
  ReSID/Musical is a source-traced auditable snapshot. Reference and control rows
  calibrate probe behavior but are not counted as named tokenizer coverage.
  Data-source labels normalize the ecosystem and subset name; each mapped-item
  count refers to that export's pinned item universe. Source evidence records
  the route used to inspect or reconstruct the mapping.}
  \label{tab:artifact-coverage}
  \small
  \setlength{\tabcolsep}{5.0pt}
  \renewcommand{\arraystretch}{1.06}
  \begin{tabularx}{\textwidth}{@{}>{\raggedright\arraybackslash}p{0.20\textwidth}>{\raggedright\arraybackslash}p{0.22\textwidth}>{\raggedright\arraybackslash}X>{\raggedleft\arraybackslash}p{0.12\textwidth}>{\centering\arraybackslash}p{0.09\textwidth}@{}}
    \toprule
    Tokenizer export & Data source & Source evidence & Mapped items & SID depth \\
    \midrule
    \multicolumn{5}{@{}l}{\textit{Source-traced named tokenizer exports}} \\
    \cmidrule(lr){1-5}
    ReSID & Amazon / Musical & Tracked snapshot & 23,742 & 3 \\
    ReSID-GAOQ & Amazon / Video Games & Code-derived export & 24,685 & 3 \\
    GRID & Amazon / Beauty (P5) & Rebuilt tokenizer & 12,101 & 3 \\
    CARD & Amazon / Beauty (P5) & Code-derived export & 12,101 & 4 \\
    DIGER & Amazon / Beauty (P5) & Checkpoint export & 12,101 & 3 \\
    DIGER & Yelp / Business & Checkpoint export & 20,033 & 3 \\
    ReSOT & Amazon / Instruments & Released archive & 6,250 & 4 \\
    LETTER & Amazon / Instruments & Released index & 9,922 & 4 \\
    LC-Rec & Amazon / Instruments & Released index & 9,922 & 4 \\
    \midrule
    \multicolumn{5}{@{}l}{\textit{Reference and control rows used for interpretation}} \\
    \cmidrule(lr){1-5}
    GRID-like & Amazon / Musical & Local stress export & 23,742 & 3 \\
    Category-prefix & Amazon / Musical & Deterministic mapping & 23,742 & 3 \\
    \bottomrule
  \end{tabularx}
\end{table}

Three comparisons support different conclusions. A same-catalog comparison
holds items, interactions, and measurement settings fixed while changing code
assignments. A cross-route profile describes heterogeneous released artifacts
without ranking tokenizer methods. A paired refresh comparison measures change
on common item IDs. Repeated depths, shards, and users are reported at their
own units; artifact and catalog dependence are addressed explicitly in the
downstream analyses. Appendix~\ref{sec:protocol-details} specifies the original
exposure protocols and admission checks.

\section{Mapping Differences Across Released Artifacts}\label{sec:mapping-differences}

\subsection{Non-Interchangeable Mapping Properties}

Table~\ref{tab:diagnostic-profile} compares addressability, prefix organization,
tail allocation, and trie structure for the eight executable source-traced
routes. Each row uses one bounded C3 execution. Same-catalog rows allow direct
comparison of their measured profiles; rows with different catalogs,
interactions, or construction settings describe the individual mappings
rather than rank tokenizer methods.

\begin{table}[!htbp]
  \centering
  \caption{D1--D5 interface profiles for eight executable source-traced
  routes. Each row uses its own catalog and reference interactions.
  D1 is the active first-level code-symbol count; D3@1 is the fraction of
  retained neighbor edges sharing a depth-1 prefix under $\pi_{\mathrm{C3}}$; D5 is the
  number of occupied full-code leaves; and D2--D4 are rates. Per-level count sequences are released
  in the deterministic table snapshot. Cross-route differences describe
  interface states under each route's declared catalog and protocol.}
  \label{tab:diagnostic-profile}
  \small
  \setlength{\tabcolsep}{3.4pt}
  \renewcommand{\arraystretch}{1.05}
  \begin{tabularx}{\textwidth}{@{}>{\raggedright\arraybackslash}p{0.255\textwidth}>{\raggedleft\arraybackslash}p{0.07\textwidth}>{\raggedleft\arraybackslash}p{0.045\textwidth}>{\raggedleft\arraybackslash}p{0.075\textwidth}>{\raggedleft\arraybackslash}p{0.085\textwidth}>{\raggedleft\arraybackslash}p{0.055\textwidth}>{\raggedleft\arraybackslash}p{0.085\textwidth}>{\raggedleft\arraybackslash}X@{}}
    \toprule
    Route & Items & \multicolumn{1}{c}{Depth} & D1 symbols & D2 collision & D3@1 & D4 tail code ratio & D5 leaves \\
    \midrule
    ReSID-GAOQ / Amazon Video & 24,685 & 3 & 32 & 0.000 & 0.213 & 1.000 & 24,685 \\
    GRID / Amazon Beauty & 12,101 & 3 & 256 & 0.205 & 0.038 & 0.957 & 10,619 \\
    CARD / Amazon Beauty & 12,101 & 4 & 256 & $<0.001$ & 0.023 & 1.000 & 12,098 \\
    DIGER / Amazon Beauty & 12,101 & 3 & 256 & 0.070 & 0.028 & 0.982 & 11,581 \\
    DIGER / Yelp Business & 20,033 & 3 & 256 & 0.102 & 0.010 & 0.941 & 18,705 \\
    ReSOT / Amazon Instruments & 6,250 & 4 & 83 & 0.000 & 0.063 & 1.000 & 6,250 \\
    LETTER / Amazon Instruments & 9,922 & 4 & 150 & 0.005 & 0.108 & 0.998 & 9,897 \\
    LC-Rec / Amazon Instruments & 9,922 & 4 & 162 & 0.005 & 0.055 & 0.998 & 9,897 \\
    \bottomrule
  \end{tabularx}
\end{table}

On the same Beauty catalog, GRID, CARD, and DIGER all use 256 first-level
symbols, yet their fractions of items sharing a full code differ substantially:
0.205, less than 0.001, and 0.070, respectively. Equal first-level code use
therefore does not imply equal item addressability. LETTER and LC-Rec likewise
have the same item count, collision rate, and tail allocation ratio, but differ
in depth-1 behavioral alignment (0.108 versus 0.055). The DIGER/Yelp and
DIGER/Beauty rows illustrate reuse of the diagnostic contract across
ecosystems. Their different checkpoints, configurations, catalogs, and
interactions prevent attributing the profile differences to the dataset alone.

LETTER and LC-Rec are recomputed under the common bounded C3 protocol rather
than the inherited CIKM settings. Appendix~\ref{sec:measurement-audit}
reconciles the exact parameter and value differences.

Figure~\ref{fig:interface-health} complements this cross-route profile with
same-catalog stress/reference and interpretation controls. Those controls test
how D2 aliasing and D3 prefix organization differ under a fixed
item universe, while Table~\ref{tab:candidate-exposure} separately bounds the
operational reach of D3.

\subsection{Allocation Within a Stratum Versus Addressability Across the Catalog}

A popularity profile can look well allocated while its items still share
addresses with the rest of the catalog. For GRID/Beauty, the tail occupies
95.7 distinct codes per 100 tail items, but only 82.2\% of tail items have a
globally unique full code. The corresponding values for DIGER/Yelp are
94.1 and 87.5\%. These are two different statistics, not two estimates of
the same collision rate. The first counts codes within a stratum; the second
counts items in singleton leaves across the catalog. GRID also contains
535 tail items whose code is shared outside the tail.

The distinction changes the follow-up decision. Allocation ratios describe
how a mapping distributes code capacity; global singleton fractions identify
which items need a resolver. A complete popularity audit should report both.
Appendix~\ref{sec:measurement-audit} supplies all eight profiles and checks the
D3 estimand against the frozen implementation.

\subsection{Interpreting a Released Mapping: ReSOT}

The released ReSOT text index for Instruments provides a concrete interpretation
case. Its 6,250 four-token codes are unique, and the normalized mapping joins
all 6,250 metadata rows and 136,226 interaction rows without missing items.
SIDScope preserves the source item IDs and pins the archive and upstream
revision. The source-only record reports derived evidence without
redistributing the archive; Appendix~\ref{sec:protocol-details} documents the
input files and reuse boundary.

\paragraph{Decision supported by the profile.}

ReSOT passes C0--C5, establishing that its profile can be reproduced from
coherent, joined inputs. The diagnostic snapshot has 6,250
unique full SIDs, zero full-code collision, per-bucket allocation ratios of
1.0, and 83, 3,645, 5,729, and 6,250 active prefixes across depths. Its edge-micro
D3 values at depths 1--3 are 0.0629, 0.0108, and 0.0043. A deterministic
control built from the same Instruments metadata and
interactions removes the shared category root, assigns the next three category
levels to prefixes, and uses an item-unique leaf. It reaches D3 values of
0.4056, 0.2094, and 0.0611 at depths 1--3 under the identical bounded protocol.
Both mappings reach zero at the item-unique fourth level. The depth-wise comparison provides the
same-dataset reference needed to interpret the released mapping.

The resource decision has two parts. First, the artifact is
\emph{eligible} for diagnostic comparison and integration testing: its addresses
are unique, its joins are complete, and its provenance is pinned. Second, its
prefix semantics require a separate check from addressability. Relative to the
same-dataset category-prefix control, the released mapping's D3 values show
weaker train-only co-occurrence recovery across depths 1--3 under the declared
protocol. A researcher can therefore proceed with a
generator while evaluating early-prefix utility as a separate design choice. The
profile supplies this distinction before final NDCG is available.

This case demonstrates the resource's basic use: establish that an export
is coherent, then compare the property relevant to the intended consumer.
Unique addresses support unambiguous resolution; the depth-wise control
separately tests the behavioral organization of the prefixes.

\section{Downstream Consequences of Mapping Structure}\label{sec:downstream}

\subsection{Construct Calibration and Its Reach}

We test whether prefixes that preserve behavioral neighbors also expose
held-out targets. For each mapping and depth, candidates share a prefix with
an item in the user's training history. Candidate recall records whether the
target enters this set, before a generator is trained or a ranker orders it.
The subsequent checks change the evaluation users, ranking rule, or candidate
pool to establish how far this association extends.

\begin{figure*}[!t]
  \centering
  \includegraphics[width=\textwidth]{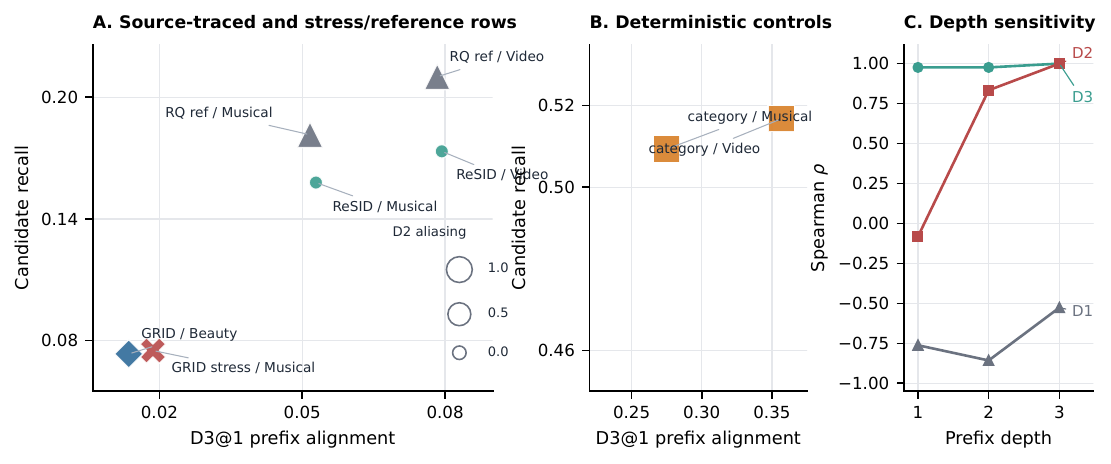}
  \caption{Mapping structure and prefix-candidate exposure. Candidate recall
  tests whether a held-out target shares a prefix with an item in the user's
  training history. Panel A expands the
  source-traced and stress/reference rows, panel B shows deterministic category
  controls on their own scale, and marker area encodes D2 full-code aliasing.
  Panel C shows that associations with candidate recall change with prefix depth.
  Category-labeled squares are interpretation controls and are excluded from
  source-traced tokenizer coverage. Labels spell out the Amazon catalog used
  by each plotted route.}
  \Description{A three-panel figure. The left panel is a scatter plot whose x-axis
  is D3 prefix alignment, y-axis is candidate recall, and marker area encodes D2
  full-code aliasing. Direct labels mark GRID/P5, ReSID, and local
  stress/reference rows. The middle panel shows deterministic category controls
  on a separate scale. The right panel plots depth-sensitive Spearman
  correlations for D1, D2, and D3.}
  \label{fig:interface-health}
\end{figure*}

Figure \ref{fig:interface-health} shows that exposure, aliasing, and depth
sensitivity move together only in part. Table
\ref{tab:candidate-exposure} orders the D3 checks by how far their endpoint
moves from the prefix-candidate protocol. Artifact--depth results describe
repeated measurements. Artifact collapse and wild-cluster analysis account
for repetition within artifacts; catalog collapse addresses shared datasets.

\begin{table}[!htbp]
  \centering
  \caption{Evidence ladder for D3 calibration reach. Block A uses the same
  prefix construction for D3 and candidate recall; Block B separates the users
  used to construct D3 from held-out exposure and ranking; Block C samples
  candidates without prefix retrieval and varies the scorer or negative pool.
  Each row reports a Spearman association with the named endpoint. Brackets are 95\% percentile bootstraps
  over the listed unit; rows labeled ``exact'' use two-sided permutation tests.}
  \label{tab:candidate-exposure}
  \small
  \setlength{\tabcolsep}{3.6pt}
  \renewcommand{\arraystretch}{1.04}
  \begin{tabularx}{\textwidth}{@{}>{\raggedright\arraybackslash}p{0.18\textwidth}>{\raggedright\arraybackslash}p{0.25\textwidth}>{\raggedright\arraybackslash}p{0.17\textwidth}>{\centering\arraybackslash}p{0.14\textwidth}Y@{}}
    \toprule
    Evaluation protocol & Measured endpoint & Reporting unit & Spearman $\rho$ & Interval / exact test \\
    \midrule
    \multicolumn{5}{l}{\textit{A. Same prefix construction}} \\
    \cmidrule(lr){1-5}
    Per-artifact aggregate & Prefix-candidate recall & 12 artifacts & 0.958 & [0.778, 1.000] \\
    Controls and local references removed & Prefix-candidate recall & 8 tokenizer exports & 0.976 & exact $p=0.0004$ \\
    Collapsed by shared catalog & Prefix-candidate recall & 5 catalogs & 0.900 & exact $p=0.083$ \\
    \midrule
    \multicolumn{5}{l}{\textit{B. Held-out users and fixed reranking}} \\
    \cmidrule(lr){1-5}
    Per-artifact aggregate & Held-out candidate recall & 8 artifacts & 0.976 & [0.793, 1.000] \\
    Prefix candidates, fixed reranker & Recall@20 & 22 artifact--depth rows & 0.645 & [0.322, 0.858] \\
    Prefix candidates, fixed reranker & NDCG@20 & 22 artifact--depth rows & 0.545 & [0.147, 0.797] \\
    \midrule
    \multicolumn{5}{l}{\textit{C. Candidates sampled without prefix retrieval}} \\
    \cmidrule(lr){1-5}
    \multirow{2}{=}{SID-prefix scorer, popular negatives} & Recall@20 & 24 artifact--depth rows & 0.817 & [0.484, 0.933] \\
    & NDCG@20 & 24 artifact--depth rows & 0.690 & [0.395, 0.898] \\
    Per-artifact aggregate & NDCG@20 & 8 artifacts & 0.690 & [-0.095, 0.988] \\
    Same scorer, popularity controlled & NDCG@20 & 24 rows; 8 artifact clusters & 0.634 & [0.115, 0.951] \\
    Metadata-category scorer & NDCG@20 & 24 artifact--depth rows & 0.207 & [-0.269, 0.588] \\
    Random-negative pool & NDCG@20 & 24 artifact--depth rows & 0.001 & [-0.346, 0.383] \\
    \bottomrule
  \end{tabularx}
\end{table}

Blocks A and B test prefix calibration because D3 and candidate exposure
consume the same prefix organization. Removing deterministic category controls
and local RQ reference rows leaves eight tokenizer-export artifacts with
$\rho=0.976$ (exact $p=0.0004$), so those controls do not create the direction.
Several exports still share catalogs and interaction logs. Collapsing them to
five catalog means gives $\rho=0.900$ (exact $p=0.083$); leave-one-catalog-out
values range from 0.800 to 1.000. The direction is stable, but five catalogs do
not support a population-level claim. With evaluation users disjoint from D3
construction, the separate eight-artifact Block B collapse remains
$\rho=0.976$ [0.793, 1.000]. Blocks B and C then move to fixed ranking,
including hard negatives constructed without prefix retrieval.

The reporting unit matters for interpreting these intervals. Artifact
clustering accounts for repeated measurements within a mapping, not for shared
catalogs; the five-catalog collapse is the cross-route sensitivity check.
Collapsing hard-negative NDCG to eight artifacts retains the point correlation
of 0.690 but widens its interval across zero [-0.095, 0.988]. Repeated
artifact--depth intervals are therefore descriptive. Appendix~\ref{sec:protocol-details}
gives the controlled regression and within-artifact robustness analysis.

\paragraph{Calibration reach.}

D3 calibrates operations that consume the prefix organization it measures.
The hard-negative SID-affinity ranker retains positive descriptive associations
with Recall@20, NDCG@20, and MRR@20, including after target-popularity
residualization. Associations weaken with co-occurrence, popularity, or
metadata-category scoring and disappear in the random-negative check.
Trie-constrained generators also use the mapping, but rank paths through
learned logits rather than the fixed affinity score. Their recommendation
quality requires a separate evaluation. Appendix~\ref{sec:nonqualifying-transfer}
reports the exploratory generator probes excluded by the prespecified
item-popularity validity check.

\subsection{Useful Prefix Assignments at a Fixed Candidate Budget}
\label{sec:fixed-budget}

Correlation with an unconstrained candidate pool leaves a practical question:
does the assignment still help when every mapping receives the same retrieval
budget? We first test GRID and CARD on their shared Beauty catalog, using a
fixed popularity reranker and exactly 100 candidate items per user. We then
predeclare the same comparison for ReSOT on the independent Instruments
catalog. Beauty supplies explicit training and validation splits. ReSOT
supplies ordered sequences but no split labels, so we exclude each user's
final event, use the penultimate event for validation, and construct histories
from earlier events. In both studies, a user hash separates the diagnostic group
from 2,000 validation users. This separates measurement users from outcome
users, not from the upstream tokenizer's training population.

For each mapping, 19 controls permute complete code assignments within
training-popularity tertiles. These controls preserve the code multiset,
prefix occupancies, and coarse popularity composition while disrupting the
assignment of items to those codes. Candidate construction takes the union of
prefix buckets reached by a user's history, excludes seen items, truncates by
training popularity, and fills any shortfall from catalog popularity. The same
popularity ranking then orders every candidate set.

\begin{table}[!htbp]
 \centering
 \caption{Fixed-budget prefix consumption on Beauty and Instruments, with
 2,000 validation users per mapping and exactly 100 candidates per row.
 Original and permuted assignments use the same candidate constructor and
 ranker; permutation values average 19 fixed controls. Catalog popularity
 retrieves without SID prefixes. Recall is a percentage; NDCG is scaled by
 $10^3$.}
 \label{tab:fixed-budget}
 \small
 \setlength{\tabcolsep}{6pt}
 \renewcommand{\arraystretch}{1.12}
 \begin{tabular*}{\textwidth}{@{\extracolsep{\fill}}lrrrrrr@{}}
 \toprule
 & \multicolumn{3}{c}{Candidate Recall@100 (\%)} & \multicolumn{3}{c}{NDCG@20 ($\times10^3$)}\\
 \cmidrule(lr){2-4}\cmidrule(lr){5-7}
 Mapping & Original & Permuted & Popularity & Original & Permuted & Popularity\\
 \midrule
 GRID / Beauty & 13.20 & 1.91 & 6.85 & 21.18 & 3.16 & 9.41 \\
CARD / Beauty & 9.70 & 1.81 & 6.85 & 15.80 & 2.83 & 9.41 \\
ReSOT / Instruments & 18.75 & 6.24 & 25.25 & 65.09 & 18.64 & 66.98 \\
 
 \bottomrule
 \end{tabular*}
\end{table}

Table~\ref{tab:fixed-budget} shows that all three original assignments expose
more held-out targets than their structure-preserving controls. The
candidate-recall gains are 11.29 percentage points for GRID (95\% paired
user-bootstrap interval [9.82, 12.80]), 7.89 points for CARD [6.63, 9.17],
and 12.51 points for ReSOT [10.81, 14.23]. NDCG also improves under the
unchanged ranker. The direction persists at budgets 50 and 200
(Appendix~\ref{sec:measurement-audit}).

The non-SID reference answers a different decision. Both Beauty mappings
outperform catalog popularity in candidate recall. On Instruments, the
assignment-specific gain over permutations coexists with higher candidate
recall from catalog popularity (25.25\% versus 18.75\%), while their NDCG@20
values are statistically indistinguishable: the paired difference is $-1.90$
[$-8.12$, 4.28] after scaling by $10^3$. ReSOT therefore carries useful prefix
signal, while the consumer choice depends on the operating objective.

Across two catalogs, the comparison establishes a concrete use for assignment
controls: they test whether a mapping's prefixes contribute beyond code
capacity and coarse popularity allocation. The popularity reference then
tests whether that contribution is sufficient for a deployment choice. The
result establishes assignment-specific utility under the tested consumers;
predicting learned-generator quality remains a separate evaluation task.

\subsection{Generated Paths and Item Resolution}

D7 follows generated SID paths back to catalog items. A trace records the
target when available, generated path, resolved items, beam rank, optional
scores, and target outcomes. Trie constraints remove invalid paths by
construction, including in vectorized production decoding
\citep{su2026static}, but valid paths can still share item addresses or repeat
items. These outcomes require separate accounting after decoding.

\paragraph{Outcome definitions.}

Target-path survival counts a target whose code appears in the beam.
\emph{Singleton credit} additionally requires that code to identify the target
alone. A resolver-specific item hit instead requires an explicit selection
rule to return the target from the leaf. The three measures share a
target-trace denominator but evaluate successive stages: path exposure,
unambiguous identification, and item selection.

The labels distinguish invalid or unresolved out-of-trie paths from outcomes
that can survive constraints: ambiguous paths, stale/out-of-catalog
resolutions, duplicate items or paths, prefix loops, and high uncertainty when
entropy is exported. Label prevalence uses beam rows; counts of affected
targets use target traces. Appendix~\ref{sec:protocol-details} specifies the
label precedence and thresholds.

\subsection{Validity Under Decoding Constraints}

We compare constrained and unconstrained decoding from the released DACT
TIGER/T5 Tools checkpoint on a fixed set of 500 targets with beam width 50. The
constrained run contains no invalid paths; the unconstrained run contains 988
invalid rows among 25,000 beams, affecting 205 targets. All 35 targets that
survive unconstrained decoding also survive constrained decoding, with four
additional constrained-only survivors: 39/500 targets versus 35/500. Recall@20
is 0.042 in both modes. The collision-free mapping makes target-path survival
and singleton credit coincide. This comparison therefore tests constraint
handling on a released checkpoint, not ambiguity prevalence.

Preparatory checks establish label coverage, scalable joining of 2,000,000
public beam rows to 240,000 target traces, and compatibility with a trained
candidate-pool export. Their protocols and outcomes are documented in
Appendix~\ref{sec:protocol-details}.

\subsection{Target Paths and Unambiguous Item Matches}

The GRID/P5 case tests valid paths that resolve to multiple items. We evaluate
the same GRU-based autoregressive architecture on two disjoint folds,
training each model on users outside its evaluation fold. We decode 500
targets per fold with beam width 50. A DIGER last-event case tests portability of the same
accounting under a different mapping and split.

Every generated path is resolved against its pinned mapping. Table~\ref{tab:g20-trained-traces}
reports ambiguous-row prevalence over beam rows, two target-level rates, and
their gap. Intervals resample users 1,000 times, retaining all targets from each
sampled user. Appendix~\ref{sec:measurement-audit} gives model, training,
decoding, and seed settings.

\begin{table}[!htbp]
  \centering
  \caption{Trained constrained-beam traces. Ambiguous-row rates use beam rows;
  target-path survival, singleton credit, and their difference use target traces.
  Singleton credit requires the target leaf to resolve to the target alone. GRID/P5
  uses two disjoint explicit-target splits; DIGER is a last-event portability
  check. All rates are percentages and the final column is a percentage-point
  difference.}
  \label{tab:g20-trained-traces}
  \small
  \setlength{\tabcolsep}{4.1pt}
  \renewcommand{\arraystretch}{1.06}
  \begin{tabularx}{\textwidth}{@{}>{\raggedright\arraybackslash}p{0.17\textwidth}>{\centering\arraybackslash}p{0.06\textwidth}>{\raggedleft\arraybackslash}p{0.075\textwidth}>{\raggedleft\arraybackslash}p{0.085\textwidth}>{\raggedleft\arraybackslash}p{0.12\textwidth}>{\raggedleft\arraybackslash}p{0.12\textwidth}>{\raggedleft\arraybackslash}p{0.12\textwidth}>{\raggedleft\arraybackslash}X@{}}
    \toprule
    Trace route & Split & Targets & Beam rows & Ambiguous rows & Path survives & Singleton credit & Gap \\
    \midrule
    GRID/P5 & 0 & 500 & 25,000 & 34.0\% & 6.4\% & 3.4\% & 3.0 pp \\
    GRID/P5 & 1 & 500 & 25,000 & 37.1\% & 5.0\% & 3.0\% & 2.0 pp \\
    DIGER sensitivity & 0 & 500 & 25,000 & 10.1\% & 6.6\% & 5.4\% & 1.2 pp \\
    \bottomrule
  \end{tabularx}
\end{table}

\begin{figure*}[t]
  \centering
  \includegraphics[width=\textwidth]{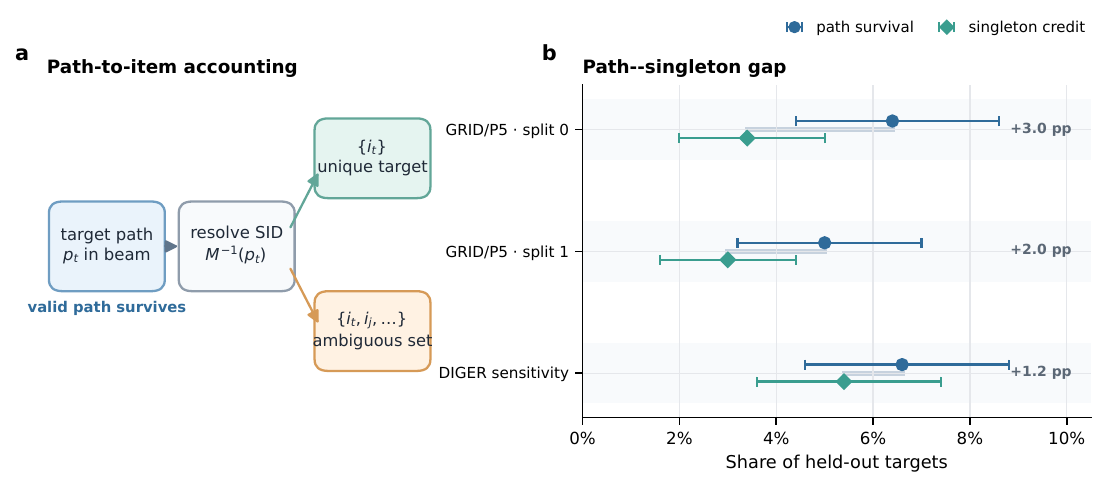}
  \caption{Path-to-item accounting in trained trie-constrained beams. (a) A
  target SID path may survive decoding yet resolve ambiguously rather than to
  the target item alone. (b) Target-path survival minus singleton credit is
  2.0--3.0 percentage points in the two GRID/P5 folds and 1.2 points in the
  DIGER sensitivity route. Points show target-level
  rates; bars are 95\% user-cluster bootstrap intervals over 1,000 resamples.
  DIGER uses a different last-event split and serves as a portability check.}
  \Description{A two-panel figure. The left panel shows a target SID path in a
  beam resolving either uniquely to the target item or ambiguously to the
  target plus other items. The right panel shows paired path-survival and
  singleton-credit rates with confidence intervals for two GRID/P5 folds and a
  DIGER sensitivity route; singleton credit is lower in all three cases.}
  \label{fig:d7-addressability-gap}
\end{figure*}

Target-path survival exceeds singleton credit by 2.0--3.0 percentage points
across the GRID/P5 splits and by 1.2 points in the DIGER sensitivity route.
The respective 95\% intervals are [4.4, 8.6]\% versus [2.0, 5.0]\%
for fold 0 and [3.2, 7.0]\% versus [1.6, 4.4]\% for fold 1.
Figure~\ref{fig:d7-addressability-gap} illustrates why this gap arises: a valid
target path can enter the beam without identifying the target alone. The
resolver comparison below tests how much of that opportunity becomes item hit.

Across all runs,
configuration, checkpoint, mapping, interaction, and labeled-trace hashes bind
the compact reported rows to the full trace exports.

The mechanism has two observable layers. Full-code aliasing in the mapping is a
necessary condition for ambiguous reverse resolution: a leaf must address more
than one item before a generated path can be ambiguous. Learned logits, decoding
constraints, and beam width then determine how often such leaves enter an
exported beam. GRID/P5 and DIGER differ in both mapping and evaluation split, and
we did not intervene on beam width while holding those factors fixed. Their
34.0--37.1\% versus 10.1\% ambiguous-row rates therefore establish sensitivity
and portability, not a causal allocation of ambiguity to the mapping or
generator.

To make the trace labels directly inspectable, the release includes 125,000
deidentified beam rows from the two GRID/P5 folds, the DIGER sensitivity case,
and constrained and unconstrained TIGER/T5 decoding. The released rows retain
the fields needed to reproduce the reported accounting---within-case trace key,
rank, decoding mode, D7 labels, resolution multiplicity, and target path/item
outcomes---while removing identity-bearing and source-sensitive fields. A
metadata record binds the five source exports by hash, and a verifier recomputes
row, trace, and label distributions.

\subsection{The Resolver Determines Which Item Credit Is Realized}
\label{sec:resolver-decision}

Singleton credit asks whether a generated leaf identifies the target
without any further choice. A deployed system can instead select one item from
an ambiguous leaf. We re-evaluate the same three frozen GRU beam exports with
four accounting rules: target-path survival, singleton credit, expected
credit under a uniform within-leaf draw, and deterministic training-popularity
resolution. Every rule uses the same path order and cutoff. The deterministic
resolver selects one item per leaf, breaking popularity ties by item ID;
it neither expands the item budget nor consults held-out targets.

The canonical reverse mapping reproduces every exported resolved-item set.
The original path and singleton outcomes also reproduce exactly.
GRID resolver counts use its declared training split. For DIGER, the last
event of each user is excluded, matching the original trace experiment.
Uniform expected credit gives a target-containing leaf of multiplicity $m$
credit $1/m$, explicitly assuming a uniform resolver. It is an expectation,
not an observed reranking outcome.

\begin{table}[!htbp]
 \centering
 \caption{Resolution policies on identical top-50 paths (500 target traces
 per case). Hit rates are percentages. Path credit is an upper reference;
 singleton credit counts targets resolved without a choice. The last column
 is the 95\% paired user-bootstrap interval for popularity resolution minus
 singleton credit, in percentage points, from 1,999 resamples.}
 \label{tab:resolver-policy}
 \small
 \setlength{\tabcolsep}{5pt}
 \renewcommand{\arraystretch}{1.1}
 \begin{tabular*}{\textwidth}{@{\extracolsep{\fill}}lrrrrr@{}}
 \toprule
 Case & Path & Singleton & Uniform & Popularity & Gain interval\\
 \midrule
 GRID / split 0 & 6.40 & 3.40 & 4.45 & 5.80 & [1.20, 3.80] \\
GRID / split 1 & 5.00 & 3.00 & 3.61 & 3.60 & [0.00, 1.40] \\
DIGER / last event & 6.60 & 5.40 & 5.82 & 6.00 & [0.00, 1.40] \\
 
 \bottomrule
 \end{tabular*}
\end{table}

For GRID split 0, popularity resolution realizes a 5.8\% hit rate from paths
with 6.4\% target survival, compared with 3.4\% singleton credit. The gain is
2.4 percentage points [1.2, 3.8]. The other GRID split and DIGER each gain
0.6 points, with intervals [0.0, 1.4]. Thus the path--singleton gap is not
automatically a realized recommendation loss: part of it can be recovered by
an explicit resolver, whose effectiveness must itself be measured.
Collision-corrected evaluation likewise distinguishes SID matching from item
credit \citep{zhang2026sidreliability}. SIDScope makes the resolution policy and
its source mapping inspectable alongside the original trace.

\section{Reuse and Adaptation After a Mapping Refresh}\label{sec:resource-use}

\subsection{Coverage Repair Changes the Model's Address Space}

A released DACT Tools mapping refresh provides a paired reuse case. The 0.6
mapping leaves 275 catalog items, including 263 interacted items, without a
SID. The 0.7 mapping closes both coverage gaps. Paired D6 re-audit records that
2,271 of 9,610 common-item codes change (23.6\% churn), while full-code
collision moves from zero to 0.000607 and depth-1 D3 moves from 0.0280 to
0.0306. The repair therefore restores addressability but also changes the
interface presented to the released generator; mapping validation alone does
not authorize checkpoint reuse.

We compare the old checkpoint,
the repaired mapping without adaptation, and three adapted models on the same
5,364 held-out users and targets. New-item recall remains zero with the repaired
mapping and old checkpoint, then reaches 12.0--12.7\% after adaptation.
The models share the same architecture.
Adaptation is ordinary full-parameter fine-tuning of the released
checkpoint on refreshed-code training sequences, with AdamW and checkpoint
selection by validation loss. The plan and runner predeclare a gate requiring recovery of at
least 90\% of the common-item NDCG loss caused by the mapping refresh, plus
nonzero recall on the 584 new-item targets. The numeric threshold is computed
from the two unadapted
states before adapted-seed outcomes are inspected. Test targets are used for neither training
nor selection. A paired user-level bootstrap, stratified into 4,780 common-item
and 584 new-item targets, quantifies uncertainty within this lifecycle case; it
was added as a reporting audit and does not alter the preregistered
point-estimate gate.

\begin{table}[!htbp]
  \centering
  \caption{DACT mapping repair and generator handoff on one fixed 5,364-target
  test universe. D6 churn is measured against mapping 0.6. The preregistered
  gate requires common-item path NDCG@20 of at least 0.01393 (13.93 in the
  displayed units) and nonzero new-item
  Recall@20. Brackets are 95\% paired user-bootstrap intervals from 4,999
  stratified resamples. Path and singleton credit coincide for every
  evaluated target in all five states. NDCG is scaled by $10^3$; new-item
  Recall is a percentage.}
  \label{tab:dact-handoff}
  \small
  \setlength{\tabcolsep}{4.0pt}
  \renewcommand{\arraystretch}{1.06}
  \begin{tabularx}{\textwidth}{@{}>{\raggedright\arraybackslash}p{0.22\textwidth}>{\raggedleft\arraybackslash}p{0.11\textwidth}>{\raggedleft\arraybackslash}p{0.11\textwidth}>{\raggedleft\arraybackslash}p{0.21\textwidth}>{\raggedleft\arraybackslash}p{0.19\textwidth}Y@{}}
    \toprule
    Mapping and model & Items without SID & Code churn vs. 0.6 & Existing-item path NDCG@20 ($\times10^3$) & New-item Recall@20 (\%) & Handoff gate \\
    \midrule
    0.6 / released model & 275 & 0.0\% & \shortstack[r]{13.94\\{[11.53, 16.50]}} & -- & Baseline \\
    0.7 / released model & 0 & 23.6\% & \shortstack[r]{13.84\\{[11.43, 16.47]}} & \shortstack[r]{0.0\\{[0.0, 0.0]}} & Fail \\
    0.7 / adapted, seed 2025 & 0 & 23.6\% & \shortstack[r]{21.00\\{[17.95, 24.12]}} & \shortstack[r]{12.0\\{[9.4, 14.7]}} & Pass \\
    0.7 / adapted, seed 2026 & 0 & 23.6\% & \shortstack[r]{20.57\\{[17.70, 23.45]}} & \shortstack[r]{12.7\\{[10.1, 15.4]}} & Pass \\
    0.7 / adapted, seed 2027 & 0 & 23.6\% & \shortstack[r]{20.26\\{[17.25, 23.24]}} & \shortstack[r]{12.7\\{[10.1, 15.4]}} & Pass \\
    \bottomrule
  \end{tabularx}
\end{table}

\subsection{A Separate Handoff Decision}

Table~\ref{tab:dact-handoff} separates coverage from model readiness. The
mapping-only state closes catalog gaps but misses the handoff gate because it
does not recover new items. Its common-item NDCG change is $-0.10$
($\times10^3$; 95\% interval $[-2.09, 1.87]$), which does not establish
ranking harm. All three adapted models improve over that state with positive
paired intervals, reach new items, and pass the complete gate in all 4,999
bootstrap resamples. This single released lifecycle case therefore supports a
concrete reuse decision; it does not isolate a churn-specific effect from the
benefit of additional training.

\section{Implications for SID Evaluation}

\subsection{Choose the Measurement for the Decision}

The three studies connect mapping inspection to different research decisions.
A tokenizer comparison needs both code allocation and global addressability:
two mappings can occupy similar numbers of codes while exposing different
items to shared addresses. A prefix consumer needs a budget-matched exposure
test, because a large candidate pool can create apparent usefulness without
meeting retrieval constraints. A generator evaluation needs an explicit
path-to-item rule, because target-path survival, singleton credit, and
resolver output measure different stages.

The fixed-budget and frozen-beam comparisons turn these measurements into
decisions: test a prefix assignment against matched controls, or choose and
evaluate a resolver for shared leaves. Each intervention holds its consumer
fixed, allowing the resulting change to be interpreted at the interface
rather than attributed to the recommendation pipeline as a whole.

\subsection{Prefix Alignment Has a Consumer-Specific Meaning}

D3 measures the preservation of retained behavioral-neighbor edges within
prefixes. Its interpretation includes the neighborhood construction, depth,
and interaction source. Construction calibration, held-out-user checks, and
non-prefix scorers establish where its connection to prefix exposure persists.

The evidence supports D3 as an operational diagnostic for prefix-consuming
retrieval and scoring. Fixed-budget controls on Beauty and Instruments show
that the released assignments expose more held-out targets than capacity- and
popularity-preserving permutations. Because these controls vary complete
assignments rather than D3 alone, they establish assignment-specific utility
under the tested consumers. Predicting learned-generator quality requires a
separate validity-passing evaluation with matched budgets and independent
catalogs as replication units.

\subsection{A Refresh Is a Joint Mapping--Model Decision}

The DACT case shows why a repaired mapping and a reusable model require
separate checks. Coverage becomes complete before the old checkpoint can
recommend new items; three fine-tuning seeds then pass the predeclared gate.
The observed recovery combines adaptation to changed codes with additional
training. An unchanged-code model receiving the same updates would distinguish
these effects; that control has not been run.

Future lifecycle records can extend the same comparison to new catalogs,
effective time windows, and repeated refreshes. Each needs stable item joins,
paired mappings, and the model state used on either side. These additions
would broaden the decision evidence while retaining the resource's central
unit: an identifiable mapping and the consumer asked to use it.

\section{Availability and Reproducibility}

\subsection{Released and Verifiable Artifacts}
The public release combines runnable code with evidence summaries. Table
\ref{tab:resource-entrypoint} separates package-verifiable surfaces from
source-bound upstream inputs. Compact snapshots, hashes, and regeneration notes
bind the reported evidence, while D7 publishes deidentified beam labels needed
to reproduce the reported accounting.

The resource entry point is the \href{https://github.com/jdding/SIDscope/releases/tag/v2.0.0}{SIDScope 2.0.0 release}.
The public package contains the original eight-table/four-figure evidence set.
The supplementary evidence archive accompanying this preprint update adds
measurement sufficient statistics and deidentified paired outcome rows for
the fixed-budget and resolver comparisons. Its CPU verifier recomputes the
reported estimates and intervals and rebuilds the four additional tables.
Raw upstream inputs remain at their sources.

The resource and repository use the \textsc{SIDScope} name; the Python import
path remains \texttt{sidinspector} for compatibility with the CIKM resource.
The manifest-approved SIDScope surface is maintained separately from the
published SIDInspector V0 repository.

The independently maintained SIDScope 2.0 software line adds a local SDK,
schema-validated generator-trace admission, protocol-bound comparisons, and
immutable evidence records. It is available through the pinned 2.0 release linked above;
the experiments in this article retain their recorded artifacts and protocols.

\begin{table}[!htbp]
  \centering
  \caption{Resource contents, supplementary evidence, and verification
  entry points. Upstream raw artifacts remain at their sources.}
  \label{tab:resource-entrypoint}
  \normalsize
  \setlength{\tabcolsep}{5.0pt}
  \renewcommand{\arraystretch}{1.06}
  \begin{tabularx}{\textwidth}{@{}>{\raggedright\arraybackslash}p{0.21\textwidth}>{\centering\arraybackslash}p{0.12\textwidth}>{\raggedright\arraybackslash}p{0.36\textwidth}Y@{}}
    \toprule
    Resource surface & Status & Evidence record & Verification gate \\
    \midrule
    Code and examples & Yes & Adapters, CLI, quickstart, failure fixture & Verify package contents \\
    Route conformance & Yes & Eight C0--C5 reports; one additional auditable snapshot & Verify route admission and evidence role \\
    Published paper evidence & Yes & Eight table snapshots, four figure records & Rebuild tables and check claims \\
    Additional analyses & Supplement & Measurement, budget, and resolver results (Tables~\ref{tab:fixed-budget}, \ref{tab:resolver-policy}, \ref{tab:measurement-audit}, and \ref{tab:budget-sensitivity}) & Recompute estimates and intervals; rebuild tables \\
    D7 trace labels & Yes & 125,000 deidentified labeled rows & Verify released trace rows \\
    Upstream raw inputs & Source-bound & Inventory, hashes, regeneration notes & Check source inventory \\
    Repository & Public & MIT license & Clone and run the verifier \\
    \bottomrule
  \end{tabularx}
\end{table}

From a clean checkout, a user can run the quickstart, rebuild all eight table
snapshots, verify route conformance and D7 labels, and rerun the invalid
fixture. The ReSOT walkthrough then demonstrates source registration,
diagnostic interpretation, and admission under the same evidence contract.

Maintenance follows the same contract. A new named route must add a route
manifest, source/license/config inventory row, C0--C5 conformance report,
compact diagnostic summary, and table/claim-ledger updates before it is counted
as admitted coverage. Backward-compatible updates may add routes, fixtures, or
documentation without changing D1--D7 semantics. Schema or diagnostic changes
preserve older table snapshots for replay. Upstream license status is rechecked
when a route changes, and routes whose redistribution
terms remain unclear continue to publish summaries and hashes rather than raw
archives.

This article substantially extends the accepted CIKM resource paper and cites
it as prior work. Table~\ref{tab:v0-delta} identifies the added artifact intake,
exposure controls, trace evidence, conformance protocol, and reproducibility
records.

\subsection{Reproducibility}

The public builder reconstructs eight table CSVs from compact released evidence
and compares every row with its frozen snapshot. Source paths, row counts,
hashes, and regeneration notes connect tables and figures to their inputs.
CPU-only checks verify package contents, sampled regeneration, archive
construction, and provenance. The supplementary archive verifies the added
tables from paired outcome rows and measurement sufficient statistics,
as distinguished in Table~\ref{tab:resource-entrypoint}. Regenerating those
outcomes from raw mappings, interactions, or beams requires the source-bound
inputs identified by the recorded hashes.

\subsection{Validated Scope and Extensions}

The evaluated mappings span Amazon and Yelp. Consumer tests cover prefix
retrieval, constrained and unconstrained beams, explicit item resolvers, and
one released DACT mapping refresh. Their comparison conditions are specified
in Sections~\ref{sec:downstream}--\ref{sec:resource-use} and the appendices.
Additional catalogs and repeated lifecycle events would broaden replication
of these decisions.

Recommended use follows the validated scope. Recommendation-quality studies
should pair the diagnostics with validity-passing trained
models. Generator-failure studies require traces with generated paths, resolved
items, beam ranks, scores, and target survival. The package supplies this
schema, representative fixtures, a public-beam accounting case, and compact
hashed summaries plus 125,000 deidentified labeled rows from the trained
GRID/P5, DIGER, and released TIGER/T5 cases.

\FloatBarrier
\section{Conclusion}

SIDScope makes Semantic-ID mappings inspectable as reusable recommendation
interfaces. Its source-traced adapters and common measurements support three
questions: how mappings differ, when those differences affect downstream
outcomes, and what must be revalidated after a refresh.

The findings connect these questions to practical choices. Reporting both
tail allocation and global singleton fractions identifies shared addresses
that a code-count profile misses. Fixed-budget controls reveal useful prefix
assignments on Beauty and Instruments, while explicit resolvers recover part
of the gap between target paths and unambiguous item matches. The released
DACT case then separates a complete mapping from a model ready to recommend
new items. Each result concerns the evaluated consumer and comparison,
providing a basis for the next test or adaptation decision.

The resource complements tokenizer and generator methods by making their
shared address space explicit. Researchers can retain mapping identity across
comparisons, select measurements that match the decision, and test a change
with a fixed consumer before attributing it to overall recommendation quality.

\begingroup
\sloppy
\hbadness=10000
\bibliographystyle{plainnat}
\bibliography{references}
\endgroup

\clearpage
\appendix
\section{Diagnostic and Evaluation Protocols}\label{sec:protocol-details}

\subsection{Prefix Exposure Protocol}
The candidate-exposure protocol uses the same normalized mapping but no
generator scores. For a user $u$, train-only history $H_u$, prefix depth $k$,
and target item $t$, the protocol forms
$C_{u,k}=\{i\in\Items\setminus H_u:\exists j\in H_u,\
\prefix_k(c(i))=\prefix_k(c(j))\}$ and records whether $t\in C_{u,k}$.
Candidate order is
deterministic: popularity is the default ranker; the fixed co-occurrence ranker
breaks score ties by training popularity and then item ID. User-level means are
accompanied by user-bootstrap intervals, while artifact-level associations
collapse repeated rows to the declared artifact or artifact--depth unit before
resampling. Candidate recall measures whether the mapping places held-out
targets in prefix regions reachable from observed history. Trained-ranker order
is evaluated separately.

\paragraph{Disjoint-user sampled ranking.}
The disjoint-user and sampled-ranking checks use a stricter protocol. Users are
assigned deterministically to four folds by a stable hash; D3 is computed from
the other three folds, while ranking metrics are computed only on the held-out
fold. The sampled protocols add each held-out target to 100 catalog negatives
after removing history items. Hard negatives are the most popular remaining
items under the diagnostic-user training slice; the falsifier samples the same
number uniformly with a seed derived from the artifact, fold, and user. The
fixed SID-affinity ranker orders candidates by the number of history items that
share the declared prefix, then by train-only co-occurrence, popularity, and
item ID. Non-prefix controls replace that first score with metadata-category,
co-occurrence, or popularity scores. We report Recall@20, NDCG@20, and MRR@20
over at most 50 held-out users per fold. Artifact--depth associations are
descriptive repeated-unit summaries; artifact-collapsed intervals provide the
effective-unit check.

\paragraph{Effective evidence unit.}
The controlled exposure model uses candidate recall as the outcome and a
standardized D3 value as the focal coefficient. It adjusts for mean log target
popularity, prefix depth, SID length, full- and prefix-collision rates,
duplicate-SID rate, and fixed indicators for dataset, artifact family, and
popularity bucket. The 1,080 rows are repeated shard--bucket observations. We
cluster by the immutable artifact key
(dataset, label, method, and manifest row) and report a 4,999-draw Rademacher
wild-cluster interval over the 12 effective artifact clusters. Artifact-level
collapse checks the effective artifact unit; the row-level fit describes
the declared conditional protocol.

The fitted D3 coefficient is 0.120 [0.067, 0.172]. With 12 artifact clusters,
this interval is a small-cluster, within-artifact robustness check. It does
not make exports sharing a catalog independent; the five-catalog collapse
in Table~\ref{tab:candidate-exposure} addresses that dependence separately.

Read together, D2 tests item addressability, D3 tests behavioral neighborhoods,
D4 tests head--tail code allocation, and D5 measures exposed trie structure. They
separate mappings that are unique but behaviorally scattered, expose useful
prefixes under concentrated branches, or alias items at full-code leaves.

\subsection{Route Inventory, Evidence Roles, and Comparison Modes}

Table \ref{tab:artifact-coverage} separates named routes from stress and
control evidence. Catalog labels preserve the upstream scope, so similarly
named routes may have different item universes. A source-traced route has a released snapshot,
upstream script path, returned archive, or reproducible generation route. A
returned archive is a run bundle recovered from an upstream or cloud pipeline
with logs, hashes, and normalized item--code outputs. Labels such as the
ReSID-GAOQ official-code-derived route describe provenance, not new \tool
methods.

Stress rows and deterministic controls instead calibrate diagnostic behavior.
The GRID-like Musical stress route, for example, contains 23,742 items but only 3,749
unique full codes, yielding a full-code aliasing rate of 0.977. It remains
separate from the source-traced GRID/P5 Beauty tokenizer-stage row. The two rows
therefore document different artifact states even though both use GRID-related
language.

These roles also govern resource extension. A new named route enters the
shared comparison set only when its mapping has a released snapshot, returned
archive, upstream script path, or reproducible tokenizer-stage route. Local
reconstructions, partial exports, and degenerate controls remain reference rows.
This preserves both comparison coverage and the evidence identity of each route.

\paragraph{Valid comparison modes.}

The resource supports three comparison modes. A same-catalog controlled
comparison holds the item universe,
interaction split, and diagnostic configuration fixed while changing the SID
assignment. This mode can attribute a difference in D1--D5 or candidate
exposure to the inspected mapping under that protocol. The Musical stress and
control rows share the catalog and interactions, making their contrast
interpretable without treating them as a tokenizer leaderboard.

A cross-route profile comparison is broader but less causal. Released indexes,
tokenizer-stage rebuilds, and official-code-derived exports may differ in
catalog, interactions, depth, and upstream objective. SIDScope can still ask
whether each artifact is addressable, how its prefixes distribute mass, and
whether its source record is reproducible. It cannot treat the resulting metric
ordering as a method ranking. This is why Table~\ref{tab:artifact-coverage}
reports catalog-specific item counts and evidence classes: each denominator belongs
to its source catalog, and the comparison concerns observable interface
properties rather than universal superiority.

A longitudinal comparison holds the source route and stable item IDs fixed
across revisions. It requires paired mappings, an explicit refresh boundary,
and old--new join coverage before interpreting D6 or changes in D1--D5. The DACT
0.6--0.7 handoff in Section~\ref{sec:resource-use} exercises this accounting on
one route; dynamic or time-conditioned artifacts should likewise enter as
versioned snapshots.

These modes determine what must accompany a reported number. Every diagnostic
record names the artifact identity, item universe, interaction split when used,
prefix depth, source role, and comparison mode. Same-catalog contrasts can
support mapping-level attribution under the fixed protocol; cross-route rows
support profile and coverage claims; longitudinal rows support change claims.
Keeping these inference units explicit prevents the resource from
turning heterogeneous public artifacts into an accidental leaderboard.

\subsection{Source, Reuse, and Admission Boundaries}

Table~\ref{tab:artifact-coverage} intentionally stops at fields needed to
compare route identity and evidence status. The released inventory binds each
route to its upstream URL and immutable revision, dataset and item count,
artifact path, SID depth and configuration, license and redistribution policy,
evidence path, hashes, and verification date. Natural display labels therefore
do not replace stable route identifiers or replay metadata.

The ReSOT source repository links a \texttt{data.zip} archive containing text
and image indexes, an item-ID sidecar, metadata, and interactions. The selected
text branch, \path{Instruments.index_lemb.json}, contains 6,250 four-token
SIDs and 6,250 unique full paths. Its inventory record binds the archive,
upstream revision, and retrieval record. The adapter translates dense integer
IDs to stable source item IDs, preserving both the metric-facing and source
identifiers. Its three normalized tables contain 6,250 mapping rows, 6,250
metadata rows, and 136,226 interaction rows, with complete item joins.
The recorded intake detected no license and therefore adopted a source-only
policy: derived summaries and hashes are included, but the archive is not
redistributed.

The inventory separates source availability from redistribution permission. A
public repository or download route makes an artifact inspectable, but it does
not by itself authorize SIDScope to republish the upstream archive. For routes
without a detected license, the package releases code, summaries, derived
tables, and hashes while requiring users to obtain upstream inputs themselves.
The GRID source uses a restricted research license, so upstream data are also
excluded. This treatment makes licensing an executable package boundary rather
than a prose footnote: the inventory verifier rejects redistribution policies
that conflict with the recorded license status.

\paragraph{Adapter admission.}

Parser success alone does not establish adapter conformance. \tool evaluates
each candidate route through six ordered contracts shown in
Table~\ref{tab:adapter-conformance}. The gates move from source identity to
schema coherence, joins, bounded diagnostics, replay identity, and finally
eligibility for the shared matrix. Controls, stress rows, and partial-coverage
routes remain separate from source-traced named coverage.

\begin{table}[!htbp]
  \centering
  \caption{Adapter admission protocol. A named route enters the shared matrix
  only when all six gates pass.}
  \label{tab:adapter-conformance}
  \small
  \setlength{\tabcolsep}{4.5pt}
  \renewcommand{\arraystretch}{1.06}
  \begin{tabularx}{\textwidth}{@{}>{\bfseries}l>{\raggedright\arraybackslash}p{0.24\textwidth}YY@{}}
    \toprule
    Gate & What must be true & Evidence checked & Failure action \\
    \midrule
    C0 & Source identity and reuse terms are explicit & URL, revision, derivation path, license, redistribution policy & Hold as discovered source \\
    C1 & SID tables are internally coherent & Required columns, level reconstruction, dataset, depth, item count & Reject normalized mapping \\
    C2 & Mapping joins metadata and interactions & Joined-item counts and missing-coverage report & Hold as partial route \\
    C3 & Bounded diagnostics execute & D1--D5 outputs with declared limits & Exclude from metric matrix \\
    C4 & Inputs are pinned and replayable & Command, runtime, evidence paths, SHA-256 hashes & Exclude from release claim \\
    C5 & Inventory identity matches the matrix role & Method, catalog, revision, config, depth, license, role & Keep as reference/control \\
    \bottomrule
  \end{tabularx}
\end{table}

The current release applies the complete C0--C5 protocol to eight heterogeneous routes
from seven method families: ReSOT's released archive, GRID's tokenizer-stage
rebuild, ReSID-GAOQ's returned mapping, LETTER and LC-Rec's released indexes,
and the CARD and DIGER code-derived exports, including DIGER checkpoints for
Beauty and Yelp. Each report records normalized row counts, bounded D1--D5
execution, three input hashes, and the exact admission identity without
redistributing raw upstream inputs. The ninth named export, ReSID/Musical,
remains a source-traced auditable snapshot rather than an executable C0--C5
route; it is counted in coverage but not among the eight conformance reports.

\subsection{Invalid-Artifact Rejection and Replay}

A public failure fixture tests whether the contract rejects plausible errors.
Its CSV schemas and item joins are valid, but one row's full \texttt{sid}
disagrees with its
\texttt{sid\_level\_*} fields. C0, C2, C3, C4, and C5 remain independently
observable, while C1 fails with the offending row index. This isolates the
rejection to inconsistent address representations, so the artifact cannot enter
the shared comparison set.

The public package supports two replay levels. With only the release checkout,
a user can validate the nine-route inventory, inspect eight frozen C0--C5
reports and their input hashes, rerun the failure fixture, and deterministically
rebuild the machine-readable and Markdown walkthrough. With the upstream ReSOT
archive supplied separately, the user can rerun normalization and bounded
diagnostics against the pinned files. The first level verifies the released
resource contract from released evidence; the second re-executes the
source-dependent intake. ReSOT is the detailed walkthrough among eight
conformant routes. This separation is the practical contribution of
the walkthrough: provenance, licensing, executable checks, diagnostic
interpretation, and route admission remain connected while upstream inputs stay
at their declared sources.

\subsection{Trace Labels and Integration Checks}

Trace flags may overlap. The primary label follows the fixed order: invalid
path, ambiguous path, stale/out-of-catalog, duplicate item, duplicate path,
prefix loop, high uncertainty, and finally valid hit. An invalid path resolves
to no catalog item; an ambiguous path resolves to more than one; a stale or
out-of-catalog unique resolution is absent from the declared active set.
Duplicate items and paths are repeats within a target trace. A prefix loop is
a repeated half-path pattern. High uncertainty requires exported entropy at
or above 2.0. Invalid and unresolved out-of-trie paths are unconstrained-only
labels; the other accounting outcomes can survive trie constraints.

Three preparatory cases check this contract. The fixture covers six
labels that can survive constraints and three invalid cases restricted to
unconstrained decoding;
\texttt{duplicate\_path} is schema-covered without a separate fixture row.
The public-beam case joins 2,000,000 valid item-expanded rows to 240,000
target traces, reproduces target-survival labels under the declared budgets,
and binds outcomes to artifact pressure, prefix load, and popularity.
The trained candidate-pool case labels all 800 rows from 40 targets
deterministically; every unique valid SID path resolves to its exported item.
These checks establish label coverage, scalable joins, and evaluation-loop
compatibility. The released T5 and controlled GRU cases in
Section~\ref{sec:downstream} supply the decoding comparisons.

\section{Measurement and Decision Checks}\label{sec:measurement-audit}

\subsection{D3 Estimands and Bounded Graph Construction}

We independently recompute all 28 available prefix-depth rows from the eight
executable routes using their pinned normalized inputs. Edge-micro recall
matches the frozen implementation exactly. Frequency-weighted recall is
\[
 D3_{\mathrm{freq}}(k;\pi)=
 \frac{\sum_{(i,j)\in E_\pi}w_{ij}
 \mathbf{1}[\prefix_k(c(i))=\prefix_k(c(j))]}
 {\sum_{(i,j)\in E_\pi}w_{ij}}.
\]
Macro recall is the mean of per-item neighbor recalls over nonempty
neighborhoods. These measure different quantities. Co-occurrence frequencies affect
neighbor selection in all three, but affect aggregation only in the
frequency-weighted version.

At depth 1, frequency weighting preserves the ordering of the eight routes
(Spearman $\rho=1.0$; maximum rank shift 0), while changing individual values
by as much as 0.0202. For example, LETTER changes from 0.107933 to 0.128145
and ReSOT from 0.062906 to 0.081391. The primary results therefore retain
edge-micro recall and name it explicitly. This ordering check does not
re-estimate the downstream evidence ladder for the weighted alternative.

\begin{table}[!htbp]
 \centering
 \caption{Measurement checks on the eight executable exports. D3 columns are
 depth-1 recalls under the same C3 graph. Tail allocation is distinct codes
 per tail item; tail singleton is the fraction of tail items whose full code
 is unique across the catalog. Tail columns are percentages.}
 \label{tab:measurement-audit}
 \small
 \setlength{\tabcolsep}{4.5pt}
 \renewcommand{\arraystretch}{1.1}
 \begin{tabular*}{\textwidth}{@{\extracolsep{\fill}}lrrrr@{}}
 \toprule
 Mapping & D3 edge-micro & D3 frequency-weighted & Tail allocation & Tail singleton\\
 \midrule
 CARD / Beauty & 0.0229 & 0.0228 & 100.0 & 100.0 \\
DIGER / Beauty & 0.0283 & 0.0284 & 98.2 & 94.2 \\
DIGER / Yelp & 0.0097 & 0.0101 & 94.1 & 87.5 \\
GRID / Beauty & 0.0384 & 0.0380 & 95.7 & 82.2 \\
LC-Rec / Instruments & 0.0550 & 0.0678 & 99.8 & 99.5 \\
LETTER / Instruments & 0.1079 & 0.1281 & 99.8 & 99.5 \\
ReSID-GAOQ / Video & 0.2128 & 0.2133 & 100.0 & 100.0 \\
ReSOT text-index / Instruments & 0.0629 & 0.0814 & 100.0 & 100.0 \\
 
 \bottomrule
 \end{tabular*}
\end{table}

LETTER and LC-Rec use the same normalized mappings, metadata, and interaction
rows as the CIKM report, apart from route labels. The earlier calculation used
20 neighbors and 1,173,333 pair events; the common C3 protocol uses five
neighbors, 9,817 pair events, and a 50-item user-eligibility threshold. Before
three-decimal rounding, LETTER changes from 0.108577 to 0.107933 and LC-Rec
from 0.052369 to 0.054972.

The original neighborhood sensitivity varied $m\in\{5,10,20\}$ and
$U\in\{50,100,200\}$ with $B=10{,}000$; all nine configurations preserved
the primary route ordering. A separate check varies the pair budget over
10,000, 100,000, and 1,000,000 and uses three seeded user orders at the
smallest budget. These choices change D3 magnitudes: across these six
configurations, GRID ranges from 0.0384 to 0.0854 and DIGER/Beauty from
0.0283 to 0.0913. Bounded pair accumulation samples part of the interaction
graph, so budget and order belong in the reported protocol. The primary route
ordering is stable over the tested $m$ and $U$ grid; reported magnitudes remain
bound to the declared budget and order.

Unit fixtures distinguish unequal co-count weights, unequal neighborhood
degrees, empty graphs, cross-stratum aliasing, and exclusion of over-length
users. D1 is also checked against joint-prefix counts: symbol marginals and
prefix counts coincide at depth 1 but measure different distributions at later
levels.

\subsection{Non-Qualifying Generator Transfer Probes}
\label{sec:nonqualifying-transfer}

Two exploratory probes asked whether D3 predicts NDCG@20 across trained
generators. The SID-code probe spans five artifacts and yields
$\rho\in[-0.564,0.205]$ with exact $p\geq0.400$; the item-ID probe spans eight
artifacts and yields $\rho=0.144$ with exact $p=0.736$. Both probes fail the
prespecified item-popularity validity check. They are therefore excluded from
the main calibration ladder and are not interpreted as positive or negative
transfer evidence. They retain the evaluated boundary: D3 calibration is
established for the tested prefix-consuming operations, while generator-quality
prediction remains a separate empirical question.

\subsection{Fixed-Budget Decision Protocol}

The fixed-budget studies use $m=5$, $U=50$, $B=100{,}000$, depth 1, 19 fixed
permutations, and candidate budgets 50, 100, and 200; budget 100 was designated
as primary before execution. SHA-256 user hashing separates diagnostic users
from 2,000 validation users. GRID and CARD are two mappings over the same
Beauty catalog, not independent catalog replications. They provide explicit
validation splits; the available LETTER inputs do not. Catalog popularity is
a secondary Beauty anchor and did not tune the prefix protocol.

The independent-catalog replication was frozen before outcome computation.
ReSOT's 136,226 Instruments events cover 6,250 mapped items and 17,112 ordered
user sequences but provide no upstream split labels. The predeclared temporal
holdout uses earlier events as history, the penultimate event as validation,
and the final event as an unused future holdout. The same partition, budgets,
ranker, and 19 controls apply to 2,000 validation users. This derived split and
the catalog-popularity reference were declared before evaluation.

The original-minus-permutation interval resamples paired validation users
1,999 times against each user's mean over 19 controls; permutation seeds are
not treated as independent tokenizers. Table~\ref{tab:budget-sensitivity}
reports all budgets. Per-user outputs retain pool size and candidate count.

\begin{table}[!htbp]
 \centering
 \caption{Budget sensitivity for the fixed prefix consumer across Beauty and
 Instruments. Candidate recall is a percentage; its gain is in percentage
 points. NDCG gains are scaled by
 $10^3$. Gains compare each original assignment with its mean permutation
 control; brackets are paired-user 95\% intervals.}
 \label{tab:budget-sensitivity}
 \small
 \setlength{\tabcolsep}{4pt}
 \renewcommand{\arraystretch}{1.12}
 \begin{tabular*}{\textwidth}{@{\extracolsep{\fill}}lrrrr@{}}
 \toprule
 Mapping & Budget & Recall & Recall gain [interval] & NDCG gain [interval]\\
 \midrule
 GRID / Beauty & 50 & 9.05 & 7.77 [6.56, 8.99] & 18.45 [14.18, 23.06] \\
GRID / Beauty & 100 & 13.20 & 11.29 [9.82, 12.80] & 18.02 [13.74, 22.65] \\
GRID / Beauty & 200 & 17.70 & 14.67 [13.00, 16.37] & 14.27 [10.67, 18.41] \\
CARD / Beauty & 50 & 7.00 & 5.81 [4.73, 6.95] & 13.39 [9.65, 17.71] \\
CARD / Beauty & 100 & 9.70 & 7.89 [6.63, 9.17] & 12.98 [9.36, 17.23] \\
CARD / Beauty & 200 & 13.20 & 9.99 [8.50, 11.43] & 10.53 [7.08, 14.43] \\
ReSOT / Instruments & 50 & 15.40 & 10.52 [9.04, 12.05] & 46.61 [38.77, 55.05] \\
ReSOT / Instruments & 100 & 18.75 & 12.51 [10.81, 14.23] & 46.44 [38.61, 54.93] \\
ReSOT / Instruments & 200 & 23.40 & 14.90 [13.10, 16.79] & 45.71 [37.71, 53.77] \\
 
 \bottomrule
 \end{tabular*}
\end{table}

\subsection{Trace and Handoff Replays}

The controlled trace generator uses a GRU history encoder and autoregressive
SID decoder with 128-dimensional item embeddings, 256-dimensional hidden
states, 12 epochs, batch size 512, dropout 0.1, learning rate $10^{-3}$, and
weight decay $10^{-4}$. Each run trains on at most 50,000 examples from users
outside the evaluation fold, using up to 50 history items and three targets
per user. Decoding normalizes each step over valid trie children and orders
beams by cumulative log probability, then lexicographic path. Resolution uses
the pinned mapping revision. User intervals use 1,000 resamples and retain all
targets of each sampled user. The base seed is 20260808 with deterministic
artifact--fold offsets.

Resolver analysis uses three source-bound GRU exports, each with 25,000 beam
rows and 500 targets. Canonical mapping reproduces path and singleton metrics
before policy comparison. Because one item is selected per leaf and distinct
leaves have disjoint item sets, the top-50 results in
Table~\ref{tab:resolver-policy} and top-20 sensitivity retain the original
path cutoff without adding candidates.

For the DACT handoff, archived outcomes are checked against source hashes.
Reviewed-manifest SHA-256 prefix \texttt{2f8f276c57f0} identifies the
predeclared plan and runner. The audit confirms that path and singleton
indicators and discounted credit coincide for every target in all five states,
preserving the gate's numerical interpretation.

\end{document}